\documentclass[11pt]{article}

\usepackage{special,times,graphicx}
\usepackage{caption}
\usepackage{float}
\usepackage{setspace}
\usepackage{amsmath}
\usepackage{amssymb}
\usepackage{authblk}
\usepackage[margin=1cm]{subcaption}
\usepackage{graphicx}
\usepackage{lineno}
\usepackage{tabularx}
\graphicspath{{.},{plots/}}

\providecommand{\keywords}[1]{\textbf{\textit{Index terms: }} #1}

\begin{document}

\title{Efficient Aerosol Retrieval for Multi-angle Imaging SpectroRadiometer (MISR): A Bayesian Approach}

\author[1]{Shijing Yao}
\author[2]{Yueqing Wang}
\author[1,2]{Bin Yu}

\affil[1]{Department of Electrical Engineering and Computer Science, University of California, Berkeley}
\affil[2]{Department of Statistics, Univeristy of California, Berkeley}

\maketitle


\begin{abstract}
Recent research in Aerosol Optical Depth (AOD) retrieval algorithms for Multi-angle Imaging SpectroRadiometer (MISR) proposed a hierarchical Bayesian model. However the inference algorithm used in their work was Markov Chain Monte Carlo (MCMC), which was reported prohibitively slow. The poor speed of MCMC dramatically limited the production feasibility of the Bayesian framework if large scale (e.g. global scale) of aerosol retrieval is desired.

In this paper, we present an alternative optimization method to mitigate the speed problem. In particular we adopt Maximize a Posteriori (MAP) approach, and apply a gradient-free ``hill-climbing" algorithm: the coordinate-wise stochastic-search. Our method has shown to be much (about 100 times) faster than MCMC, easier to converge, and insensitive to hyper parameters. To further scale our approach, we parallelized our method using Apache Spark, which achieves linear speed-up w.r.t number of CPU cores up to 16. Due to these efforts, we are able to retrieve AOD at much finer resolution (1.1km) with a tiny fraction of time consumption compared with existing methods.

During our research, we find that in low AOD levels, the Bayesian network tends to produce overestimated retrievals. We also find that high absorbing aerosol types are retrieved at the same time. This is likely caused by the Dirichlet prior for aerosol types, as it is shown to encourage selecting absorbing types in practice. After changing Dirichlet to uniform, the AOD retrievals show excellent agreement with ground measurement in all levels.

\keywords{MISR, remote sensing, high resolution, aerosol retrieval, Bayesian, Markov Chain Monte Carlo, Maximize a Posteriori, stochastic search, coordinate descent}

\end{abstract}

\section{Introduction}
Atmospheric aerosols are a colloid of fine particles suspended in the air. They can be solid particles or liquid droplet, with sizes ranging from 0.01 $\mu m$ to around 50 $\mu m$. Aerosols occur mostly from natural processes (e.g. volcanoes, dust storms, forest fires, sea spray and etc.), but can also come from human activities (e.g. burning of fossil fuels). Despite their small overall proportion, artificial aerosols can dominate the atmosphere above urban and industrial regions, which can significantly affect human health and life expectancy \cite{Ulrich}. Aerosols with diameter less than 2.5 $\mu m$, also known as $PM_{2.5}$, are especially harmful to human respiratory system, as they penetrate into the gas exchange regions of the lung and may pass through the lungs to affect other organs.

Air pollution associated with excessive aerosols obscures the air, and severely reduces visibility \cite{martin}. The (column) amount of atmospheric aerosols is characterized by Aerosol Optical Depth (AOD). It is defined as the accumulated radiative energy loss due to aerosol scattering and absorbing in the sunlight’s pathway. AOD can be retrieved by measuring the solar reflectance from aerosols. The Multi-angle Imaging SpectroRadiometer (MISR) is a set of satellite instrument that measures such reflectance. It was launched by NASA, aboard Spacecraft Terra, and is the first multi-angle and multi-spectral satellite radiometer orbiting the Earth. MISR is uniquely designed for AOD retrievals, as its multi-angle imaging provides strong visual contrast of aerosols \cite{diner1998}. The wide range of its spectral coverage also exploits the wavelength dependency of aerosol opacity, which can be used to estimate the size distribution of the aerosol particulates.

The NASA operational algorithm for MISR AOD retrieval is based on an error minimization between modeled solar reflectance and MISR observation \cite{MISRAlgTheory,martonchik1998a,martonchik1998b}. The algorithm is able to inversely infer AOD and aerosol composition if the modeled top-of-atmosphere reflection matches with MISR observation. To find the correct solutions, NASA operational algorithm searches for aerosol composition from a pre-defined subset instead of from the complete $(M-1)-$simplex parameter space, $M$ being the total number of possible aerosol types. In the AOD dimension, the algorithm searches along a one-dimensional grid, with AOD values ranging from 0 to 6.

This retrieval scheme is performed independently region by region, where each region is defined as a 17.6km $\times$ 17.6km grid. In urban areas however, complex anthropogenic aerosol sources (e.g. vehicles, power plants, and factories that burn fossil fuels) could have high spatial heterogeneity even within one region, thus profiling spatial distribution of aerosols at finer resolution ($<$ 17.6~km) is very much desirable. At smaller grid size, input image data becomes spatially noisy and so do independent grid-wise retrievals.

To incorporate spatial smoothness for AOD and allow aerosol composition to be selected in the whole $(M-1)-$simplex, a hierarchical Bayesian model was proposed \cite{WanJiaYuJia12}. A case study by Moon et.al. \cite{taesup} shows the AOD retrieval based on Bayesian method has better accuracy compared with ground data than that based on the NASA operational algorithm. The ground data they used was AErosol RObotic NETwork (AERONET) Distributed Regional Aerosol Gridded Observation Networks (DRAGON) campaign data, which was obtained in the summer of 2011. Despite of these advantages, the use of MCMC in the hierarchical Bayesian model makes the algorithm prohibitively slow. It was also not well understood why the Bayesian model severely overestimates AOD on low-AOD regions.

To improve the computational speed, we propose using MAP: a much more efficient method to make inference of the Bayesian model. We choose coordinate-wise stochastic search to do hill-climbing. It runs hundreds of times faster than the previous MCMC method while still achieves superior accuracy. Parallelization of these methods for further speed improvement is also developed. Regarding the overestimation observed in \cite{taesup}, we find that the error is caused by erroneous aerosol component retrieval.

The rest of the paper is organized as follows; in Section \ref{sec:review}, we briefly review the mathematics of NASA operational algorithm and the existing hierarchical Bayesian model. In Section \ref{sec:algorithm}, we describe how MAP is applied, using coordinate-wise stochastic search. We also show how to incorporate parallelization to MAP. Section \ref{sec:results} compares the performance of MAP and MCMC in terms of speed and accuracy. In Section \ref{sec:discussion}, we give a thorough investigation on the root case of overestimation in existing work \cite{taesup}, and how the Bayesian model gives superior performance in presence of noisy data. We also show that our method can provide uncertainty measure for the AOD retrievals. Section \ref{sec:conclusion} concludes the paper with a discussion on possible future research directions.

\section{Mathematics of MISR AOD Retrieval}
\label{sec:review}
\subsection{NASA Operational Algorithm}
MISR observes solar reflectance at the top of atmosphere from 9 different angles at 4 different spectral bands (wavelengths: 446, 558, 672, and 866 nm). The camera viewing angles are pointed in the nadir, forward, and aft directions ($0^0$,$\pm 26.1^0$,$\pm 45.6^0$,$\pm 60.0^0$,$\pm 70.5^0$ w.r.t nadir). MISR collects a swath of imagery only on the daylit side of the Earth. Each orbit is then divided into 180 blocks, with each block having a size of 563.2~km (cross-track) $\times$ 140.8~km (along-track) in global mode. Each block is composed of arrays of regions, with each region having a grid size of 17.6~km $\times$ 17.6~km. The NASA operational algorithms are applied grid-by-grid independently to retrieve AOD and aerosol composition information.

There are 36 (4$\times$9) total channels of image data for each region $p$. Let $\overrightarrow{L_p}=[L_{1,p},L_{2,p},...,L_{36,p}]$ where $\overrightarrow{L_p}$ represents reflected radiance in all 36 channels at region $p$, and each component $L_{c,p}$ ($c=\{1,2,...,36\}$) represents reflectance in the $c$th channel at region $p$. 

The NASA operational algorithm models $\overrightarrow{L_p}$ by solving nontrivial partial differential equations (PDE) based on the scattering physics between solar radiance and aerosols, given proper boundary conditions at Earth surface and top of atmosphere \cite{misr_ancillary,chandrasekhar,diner1984a,diner1984b}. The solution of the PDE at region $p$ is has the form $\overrightarrow{L^{RT}}(\tau_p,\overrightarrow{\theta_p})$, which maps AOD $\tau_p$, aerosol composition $\overrightarrow{\theta_p}$ to all 36 channels of reflected radiance at region $p$. The composition vector $\overrightarrow{\theta_p}$ is defined on $(M-1)-$simplex, where $M$ is the total number of possible aerosol types. $\overrightarrow{\theta_p}$ is also called aerosol mixing percentages or mixing vector in different literatures. In MISR aerosol retrieval, the modeled solar reflectance $\overrightarrow{L^{RT}}(\tau_p,\overrightarrow{\theta_p})$ is compared against the actual MISR observation $\overrightarrow{L_p}$ to determine whether the values of $(\tau_p,\overrightarrow{\theta_p})$ is appropriate. The algorithm retrieves $(\tau_p,\overrightarrow{\theta_p})$ pair, by finding the one that makes $\overrightarrow{L^{RT}}(\tau_p,\overrightarrow{\theta_p})$ close enough to $\overrightarrow{L_p}$. The closeness measure is in the form of weighted least square error:

 \begin{linenomath*}
     \begin{equation}
     \chi_p^2=\sum_{c=1}^{36}\frac{(L_{p,c}-L_{c}^{\text{RT}}(\tau_p,\overrightarrow{\theta_p}))^2}{2\sigma^2_c}\tag{Eq.1}\label{eq:chi_square}
     \end{equation}
    \end{linenomath*}
    
where $\sigma_c^2$ is the estimated variance of in channel $c$. The algorithm determines the AOD retrieval in region $p$ is successful if it can find at least one pair of $(\tau_p,\overrightarrow{\theta_p})$ that makes (\ref{eq:chi_square}) smaller than some pre-defined threshold value, and unsuccessful if otherwise. When multiple pairs of $(\tau_p,\overrightarrow{\theta_p})$ end up lower than the predefined error, their mean value is used as the retrieval.

\begin{table*}[h]
    \caption{MISR Version 22 aerosol components and their characteristics from \cite{kahn2015}}
    \begin{center}
        \begin{tabular}{ p{0.05\textwidth} | p{0.3\textwidth} || p{0.08\textwidth}  | p{0.08\textwidth} | p{0.1\textwidth} | p{0.1\textwidth} | p{0.1\textwidth}}
            \hline
            \scriptsize{\textbf{No.}} & \scriptsize{\textbf{Category}} & \scriptsize{\textbf{Min}} $r_{\text{min}}(\mu m)$  &  \scriptsize{\textbf{Max}} $r_{\text{max}}(\mu m)$ & \scriptsize{\textbf{Characteristic Radius}} $r_{\text{c}} (\mu m)$ & \scriptsize{\textbf{Distribution Width}} $\sigma$ & \scriptsize{\textbf{SSA} (558nm)}  \\
            \hline
            \hline
            1 & small\_spherical\_nonabsorb& 0.0010 & 0.4000 & 0.03 &1.65 & 1.00\\
            2 & small\_spherical\_nonabsorb& 0.0010 & 0.7500 & 0.06 &1.70 & 1.00\\
            3 & medium\_spherical\_nonabsorb& 0.0010 & 1.5000 & 0.12 &1.75 & 1.00\\
            6 & large\_spherical\_nonabsorb& 0.1000 & 50.000 & 1.00 &1.90 & 1.00\\
            8 & small\_spherical\_moderate\_absorb& 0.0010 & 0.7500 & 0.06 &1.70 & 0.90\\
            14 & small\_spherical\_strong\_absorb& 0.0010 & 0.7500 & 0.06 &1.70 & 0.80\\
            19 & medium\_dust& 0.1000 & 1.0000 & 0.50 &1.50 & 0.98\\
            21 & coarse\_dust& 0.1000 & 6.0000 & 1.00 &2.00 & 0.90\\
            \hline
        \end{tabular}
    \end{center}
    \label{tab:misr_comp}
\end{table*}

For NASA operational algorithm, $\tau_p$ is referenced to green band. Each candidate aerosol composition $\overrightarrow{\theta_p}$ has up to 3 components (total number of possible aerosol types is eight). The percentages of the non-selected components in a composition vector are set zero. Table \ref{tab:misr_comp} shows some physical properties for these eight aerosol types. Note that component indices are $\{1,2,3,6,8,14,19,21\}$ because only these components are most common and the rest are simply neglected.

One limitation of the NASA operational algorithm is that only 13 AOD levels and 74 pre-defined aerosol compositions are considered in the search space of $(\tau_p,\overrightarrow{\theta_p})$. When more than 3 aerosols are present, or the aerosol-type combination is not included in the pre-defined composition, or the aerosol mixing percentages are very different from the pre-defined values, the limited search space could potentially make the retrieval algorithm sub-optimal or even fail.

Another limitation is that in urban areas where aerosol spatial distribution tends to have large heterogeneity, aerosol retrievals may need to be done on a finer grid ($<17.6$~km). As the resolution increases, the input MISR image is noisier due to less spatial averaging. The grid-wise operational algorithm could overfit the observed data.

\subsection{Hierarchical Bayesian Model}
To address the issues in the operational algorithms, Wang et.al. \cite{WanJiaYuJia12} proposed a hierarchical Bayesian model. A spatial smoothness model for AOD, characterized by an intrinsic Gaussian-Markov random field (GMRF) and a flexible continuous model for aerosol composition, characterized by a Dirichlet distribution was introduced. The complete joint likelihood of the whole model is described in Equation \eqref{equ:aod_obj}.

\begin{figure}[h]
    \centering
    \includegraphics[width=0.4\textwidth]{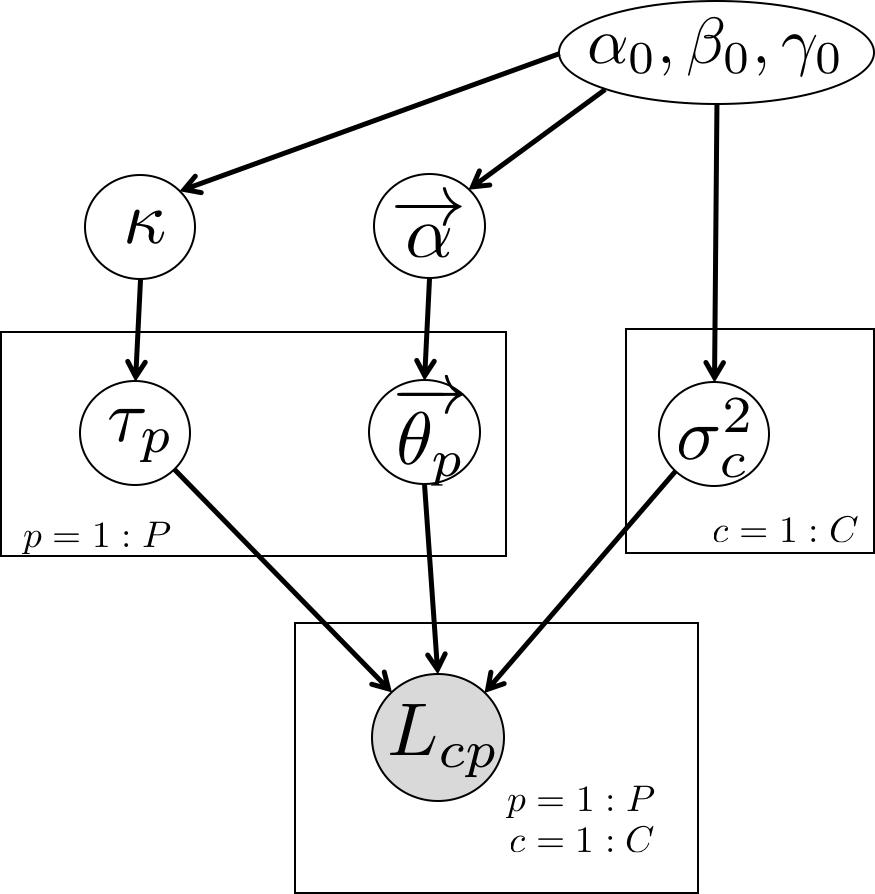}
    \caption{Hierarchical Bayesian Model}
    \label{fig:bayesian}
\end{figure}

\begin{multline}
\label{equ:aod_obj}
f(\overrightarrow{\tau},\overrightarrow{\theta},\overrightarrow{\sigma^2},\kappa,\overrightarrow{\alpha}|\overrightarrow{L})  \propto \frac{\kappa^{\frac{P-3}{2}}\Gamma(\sum\limits_{m=1}^M\alpha_m)}{\prod\limits_{c=1}^C(\sqrt{2\pi}\sigma_c)^{P+2}\prod\limits_{m=1}^M\Gamma(\alpha_m)}
\exp \Bigg\{  -\sum\limits_{p=1}^{P}\sum\limits_{c=1}^{C}\frac{\left(L_{cp}-  L_{c}^{RT}\left(\tau_p,\overrightarrow{\theta_p}\right)\right)^2}{2\sigma_c^2} \\
-\frac{1}{2}\kappa\sum\limits_{p^{\prime}\sim p}(\tau_{p^\prime}-\tau_p)^2 + \sum\limits_{p=1}^{P}\sum\limits_{m=1}^{M}(\alpha_m-1)\log\theta_{pm} \Bigg\}
\end{multline}

$\kappa$ characterizes the spatial smoothness of the GMRF, $\alpha$ characterizes the aerosol composition sparsity, $p^\prime$ is the index of every immediate neighbors of region $p$. $C=36$ if all channels are available, $P$ is usually around $1000 \sim 2000$, and $M=8$.

Wang et.al. \cite{WanJiaYuJia12} used MCMC to draw posterior samples of $\tau_p$ and $\overrightarrow{\theta_p}$, $p=1\ \text{to} \ P$. The computational speed is exceedingly slow which is the primary motivation of this work.
\section{Inference Algorithms}
\label{sec:algorithm}

Unlike using MCMC in the previous work, we use MAP to make the inference much faster. In particular we choose coordinate-wise stochastic search for the hill climbing. The method is inspired from Metropolis-Hastings (M-H) update, but converts M-H update into a greedy one - always reject a new sample if the posterior probability is not improved. The stochastic jump in each update gives the algorithm the potential to escape from local optima. A specially designed proposal distribution could regularize the variables (e.g. encourage sparsity for $\overrightarrow{\theta_p}$), which plain coordinate descent method cannot achieve.

\subsection{Coordinate-wise Stochastic Search}

Table \ref{tab:randomized_local_search} shows a skeleton of coordinate-wise stochastic search.

\begin{table}[H]
    \caption{Speed-up Inference Algorithm: Coordinate-wise Stochastic Search}
    \label{tab:randomized_local_search}
    \centering
    \begin{tabular}{|l|}
        \hline
        \textbf{initialize} $\tau_{1:P}$,$\theta_{1:P}$ \\
        \hline
        \textbf{while} $True$ \\
        \hspace{4pt}\textbf{for} $p=1$ to $P$ \\
        \hspace{8pt}Propose $\tau^\star_p \sim q(\tau_p|\tau_{-p}) \propto \exp\left\{\ -\frac{1}{2\Delta^2}\left( \tau_p - \frac{1}{n_p}\sum\limits_{p\prime \sim p}\tau_{p\prime}   \right)^2\right\}$\\
        \hspace{8pt}Update $\tau_p \leftarrow \tau^\star_p$ if $f(\sigma^2,\tau^\star_p,\theta,\kappa|L) - f(\sigma^2,\tau_p,\theta,\kappa|L) > 0$\\
        \hspace{8pt}Update $\kappa \leftarrow \frac{P-3}{\sum\limits_{p\prime \sim p} (\tau_p-\tau_{p\prime})^2}$\\
        \hspace{8pt}Propose $\theta^\star_p \sim q(\theta_p|\theta_{-p}): \theta_{p,k} \sim Gamma\left( \frac{1}{n_p}\sum\limits_{p\prime \sim p} \theta_{p\prime,k},1\right)$ \\
        \hspace{8pt}$\theta_{p,k} = \frac{\theta_{p,k}}{\sum\limits_{k=1}^K \theta_{p,k}}$, $k$ = $1$ to $K$\\
        \hspace{8pt}Update $\theta_p \leftarrow \theta^\star_p$ if $f(\sigma^2,\tau,\theta^\star_p,\kappa|L) - f(\sigma^2,\tau,\theta_p,\kappa|L) > 0$\\
        \hspace{8pt}Update $\sigma_c^2 \leftarrow \frac{\sum\limits_{p=1}^{P}\left(L_{cp}-L_c^{RT}(\tau_p,\theta_p)\right)^2}{P+2}$, $c$ = $1$ to $C$\\
        \hspace{4pt}\textbf{end for} \\
        \hspace{4pt}\textbf{break}, if $|\Delta f(\sigma^2,\tau,\theta,\kappa|L)| < \epsilon$\\
        \textbf{end while}\\
        \hline
        \textbf{return} $\tau$,\ $\theta$,\ $\sigma^2$\\
        \hline
    \end{tabular}
\end{table}

\subsection{Parallelization in Spark}
Our MAP method needs to make inference in high dimensions. $(\tau_p,\overrightarrow{\theta_p})$, $p=1 \sim P$ ($P \approx 18000$) are updated in a cyclic manner: from the beginning to the end of an image block. This renders repeatedly usage of sequential loops and dramatically hurts the speed of our algorithm. On the other hand, the relative independence between distant pixels within a MISR block admits a natural fit with parallel computing. To take advantage of the relative weak coupling between pixels that are far apart, we have developed a parallelized parameter updating scheme. 

Suppose $n$ CPU cores are available, one block is partitioned into $n$ non-overlapping patches. In each patch, parameters are updated sequentially from the first pixel to the last. However computation in different patches are carried out simultaneously. When $n$ is equal to the number of pixels in an image block, the parallelization reaches its upper limit of speed-up.

A careful reader may find that our algorithm requires neighborhood information when updating $\tau_p$ and $\overrightarrow{\theta_p}$ while current neighborhood information may not be available in the parallel setting. To mitigate this problem, we keep a separate copy of the last iteration of all variables, and use those values of the neighborhood as a surrogate.

We implement the above idea using Apache Spark \cite{spark} to scale the parallelism up to 64 cores. The speed improvement due to parallelization is shown in \ref{fig:spark_scaling}.

\begin{figure}[H]
	\centering
	\includegraphics[width=0.8\textwidth]{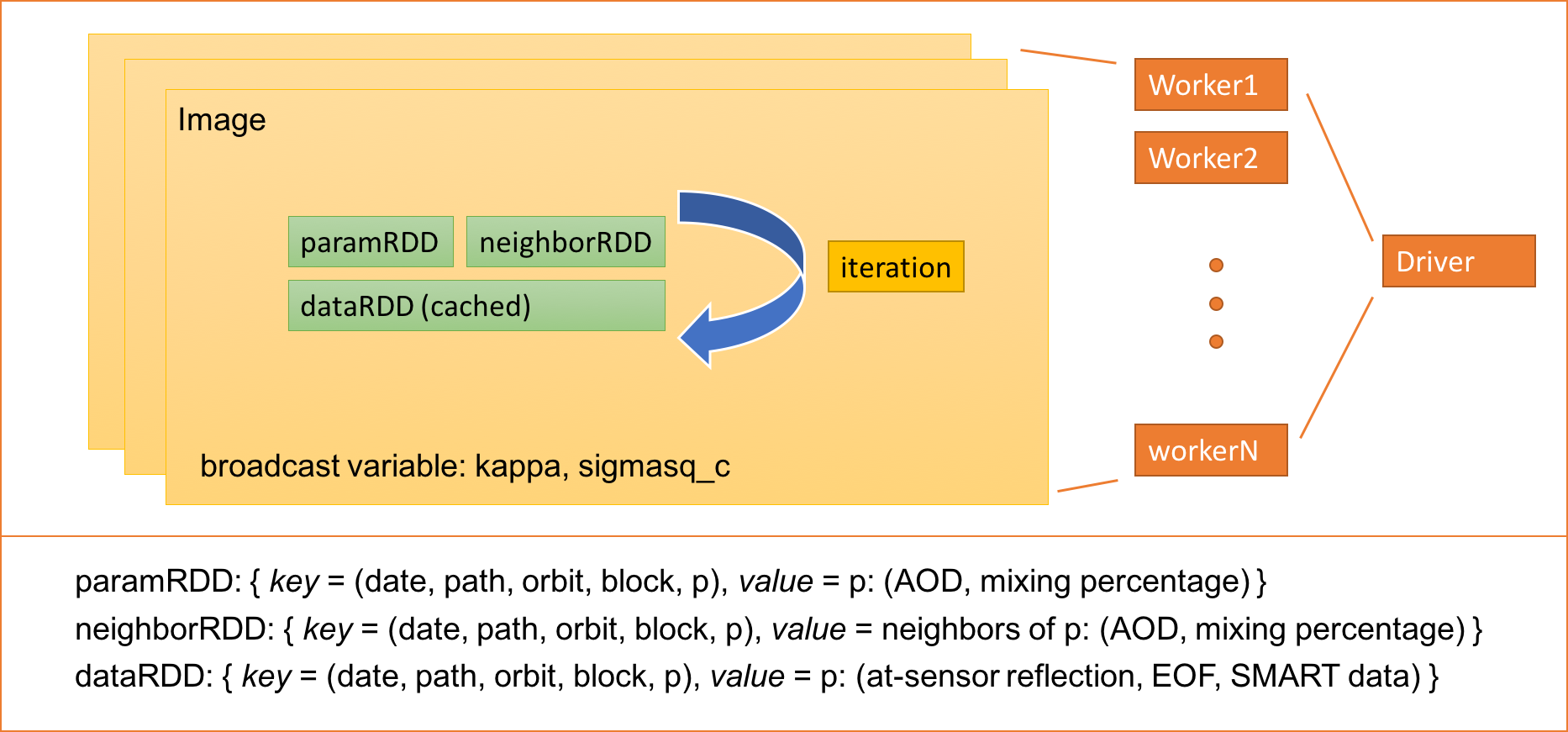}
	\caption{Algorithm parallelization implemented in Spark. Image information is stored as Resilient Distributed Data (RDD) to enable parallelization among patches.}
	\label{fig:spark}
\end{figure}

Apache Spark has been widely recognized and adopted as a fast and general engine for large-scale data processing. It has an advanced DAG execution engine that supports cyclic data flow and in-memory computing. Spark revolves around the concept of a resilient distributed dataset (RDD), which is a fault-tolerant collection of elements that can be operated on in parallel. Our AOD retrieval algorithm is admittedly iterative, but still parallelizable. We hope the patch-wise update can take advantage of the spark capabilities. Figure \ref{fig:spark} demonstrate the highlight of our implementation in the spark system. We find that our algorithm scales up reasonably well as we increase the number of cores in the computation.

We use Pyspark as the API for Spark, which is compatible with MISR Science Team. To insure efficiency for computationally expensive jobs, we use C code for the core retrieval subroutines, such as the interpolation operations in SMART (Simulated MISR Atmospheric Radiative Transfer) data \cite{misr_ancillary}. In terms of hardware, we use Amazon EC2 instance (c3.8xlarge) to do our experiment. c3.8xlarge has 32 cores and 60G memory. The CPU is Intel Xeon E5-2670 with 2.8G clock frequency, and its networking performance is 10Gigabit. We run our spark program both in local mode and Hadoop cluster mode. Local mode scales up slightly better than cluster mode due to less network overhead for broadcast variables.

\section{Results}
\label{sec:results}

\begin{figure}[H]
	\centering
	\includegraphics[width=.6\textwidth]{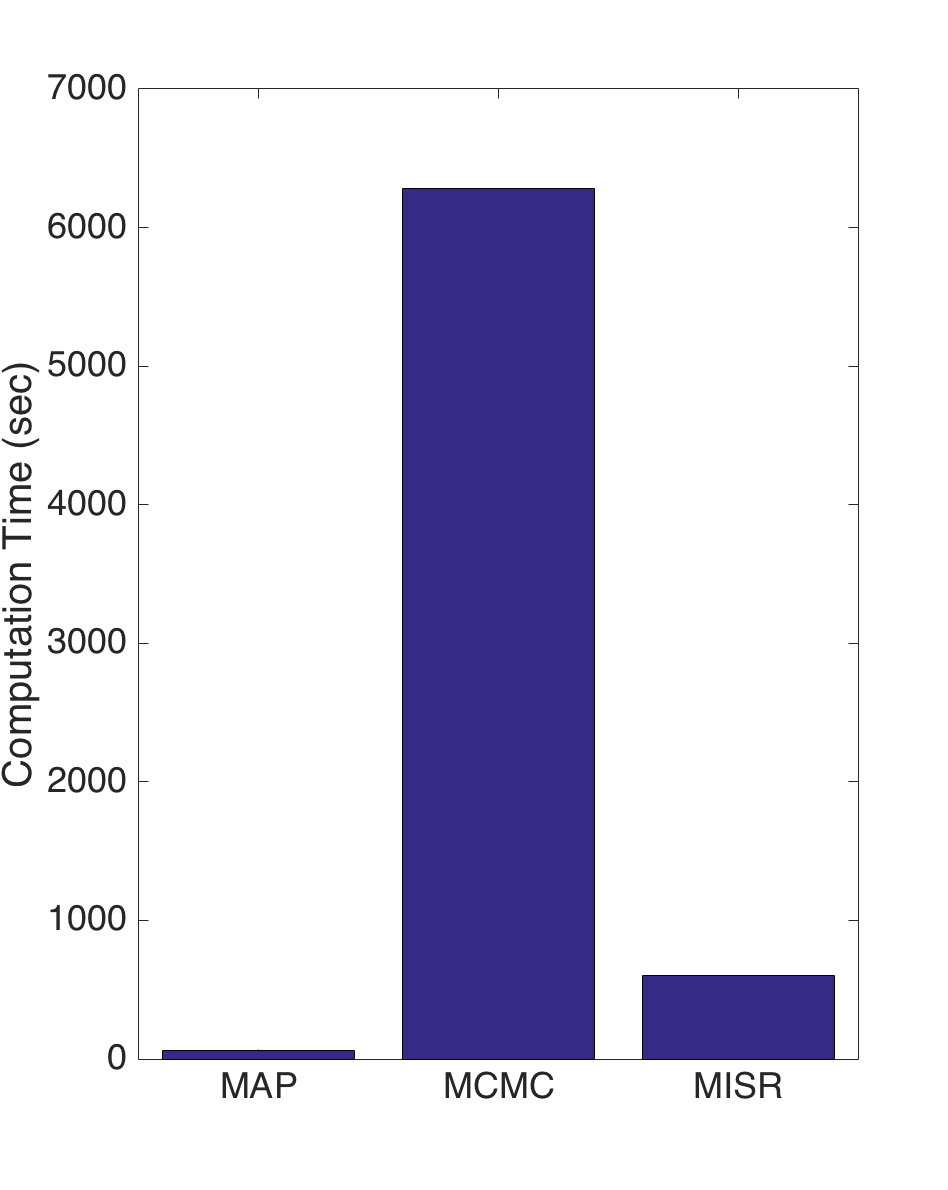}
	\caption{Speed performance: MAP, MCMC and MISR. MAP is the fastest among the three.}
	\label{fig:speed}
\end{figure}

Since hill-climbing algorithms achieve objective function improvement at every iteration, the convergence speed of MAP is much faster than MCMC. In practice it is also much easier to detect their convergence compared to MCMC. Our experiment is carried out on MacOS with 2.3G Intel Core i7 and Linux with 2.6G AMD Opteron 8384. On both platforms, MAP is about 100 times faster than MCMC. Given the same spatial granularity (4.4 km), Figure \ref{fig:speed} shows that MAP spends only a tiny fraction of the time spent by MCMC. It's also faster than MISR operational method.

We also use the AERONET data for accuracy evaluation, same as Moon et.al. \cite{taesup}. We list the data information again in Table \ref{tab:misr_overpass}. 

\begin{table}[H] 
    \caption{MISR overpass times for the Baltimore-Washington region.}
    \begin{center}
        \begin{tabular}{ c ||  c  | c | c }
            \hline
            \textbf{Date}  &\textbf{Path number}  &  \textbf{Orbit number} & \textbf{Overpass time}     \\
            \hline
            \hline
            June 2, 2011 & 16 & 60934 & 16:03  \\
            June 4, 2011 & 14 & 60963 & 15:51  \\
            July 20, 2011 & 16 & 61633 & 16:03  \\
            July 22, 2011 & 14 & 61662 & 15:51 \\
            July 29, 2011 & 15 & 61764 & 15:57 \\
            \hline
        \end{tabular}
    \end{center}\vspace{-.2in} \label{tab:misr_overpass}
\end{table}

\begin{figure}[H]
    \centering
    \includegraphics[width=.8\textwidth]{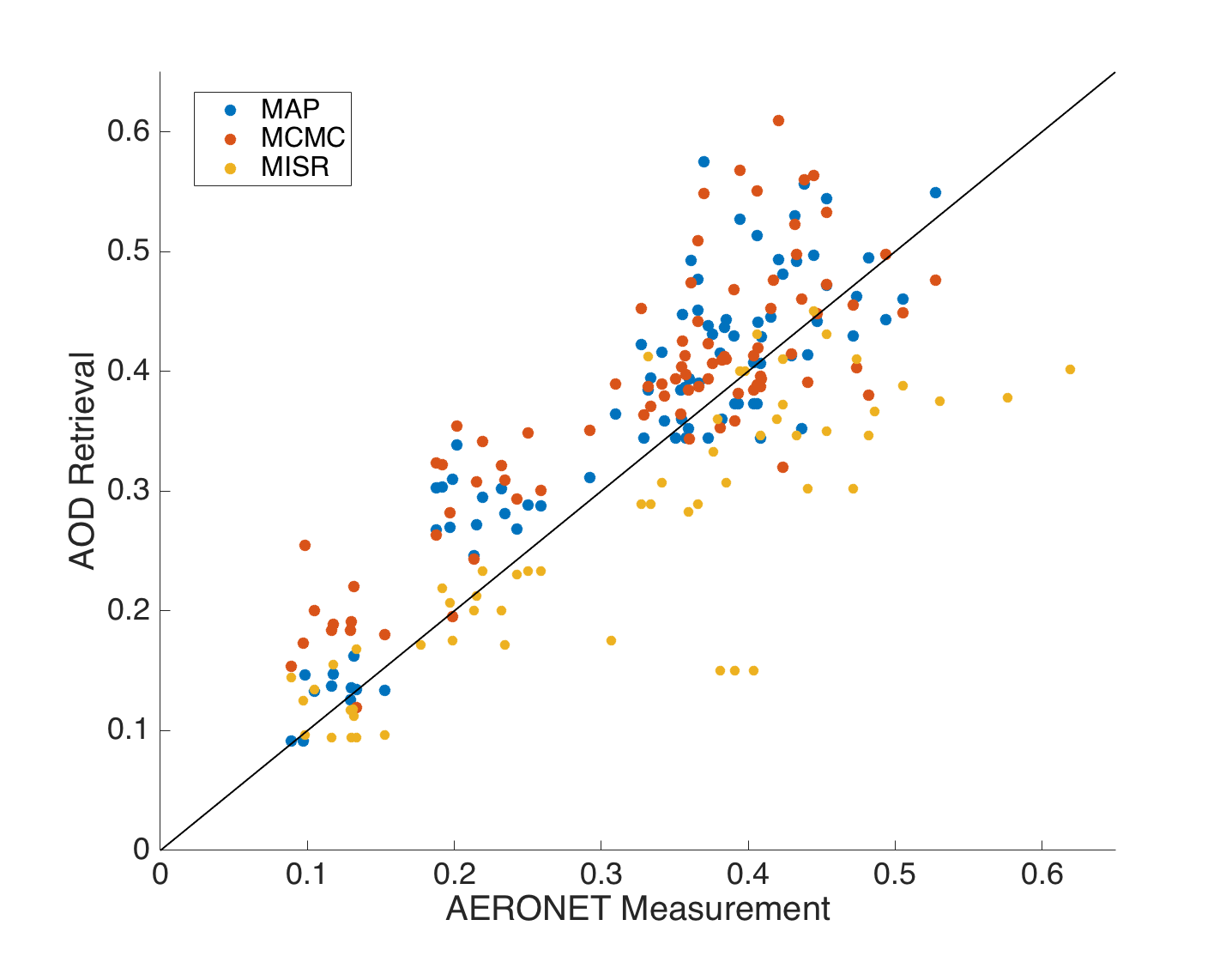}
    \caption{Retrieved AOD v.s. ground measurement given the same spatial granularity (4.4km). MAP has the best accuracy in all AOD levels.}
    \label{fig:scatter}
\end{figure}

Figure \ref{fig:scatter} visually compares MAP, MCMC and MISR when compared with ground measurement with the same 4.4km grid size. Table \ref{tab:performance} has more details. Both MAP and MCMC outperform MISR operational algorithm significantly. For medium or high AOD levels, MISR operational algorithm underestimates aerosols significantly while Bayesian model agrees with AERONET measurement very well. For low AOD levels, MCMC overestimates aerosols, as reported in \cite{taesup} while MAP is able to track ground measurement reasonably well. A detailed study in Section \ref{sec:discussion} will address this phenomenon.

\begin{table}[H]
	\caption{Performance Comparison with the Same Grid Size (4.4km)}
	\label{tab:performance}
	\begin{center}
		\begin{tabular}{|c|c|c|c|c|c|}
			\hline
			& MAP & MCMC\\
			\hline
			\hline
			Time (sec) & 60 &  6250 \\
			\hline
			Iteration \# & 20 & 1000 \\
			\hline
			Correlation & 0.939 & 0.862 \\
			\hline
			RMSE & 0.042 & 0.074 \\
			\hline
			Mean Bias & 0.016  & 0.047 \\
			\hline
		\end{tabular}
	\end{center}
\end{table}

\begin{figure}
    \centering
    \begin{subfigure}{0.8\textwidth}
        \centering
        \includegraphics[width=\textwidth]{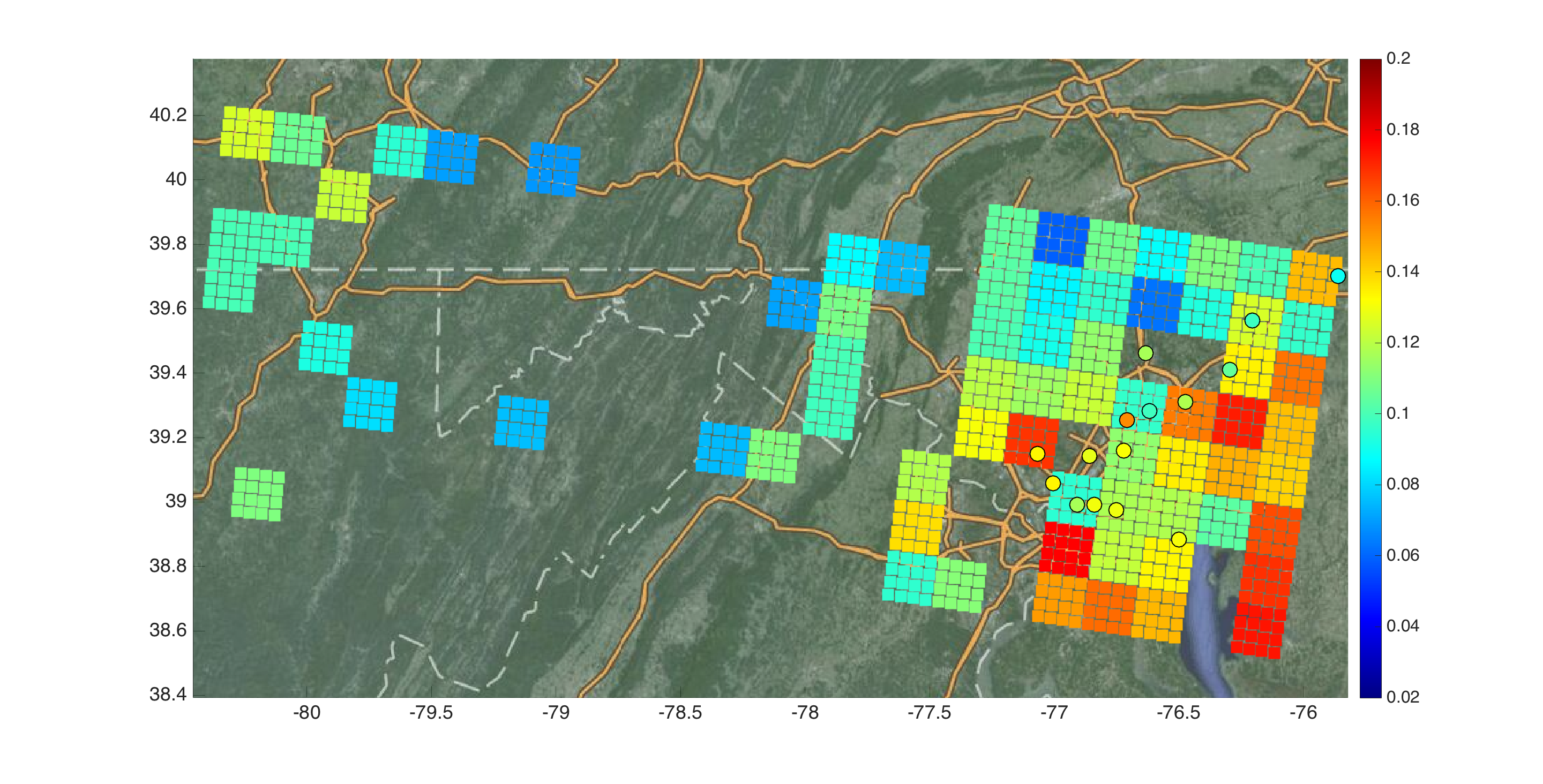}
        \caption{MISR Operational Algorithm: $17.6$km $\times$ $17.6$km}
        \label{fig:overlay1}
    \end{subfigure}
    \begin{subfigure}{0.8\textwidth}
        \centering
        \includegraphics[width=\textwidth]{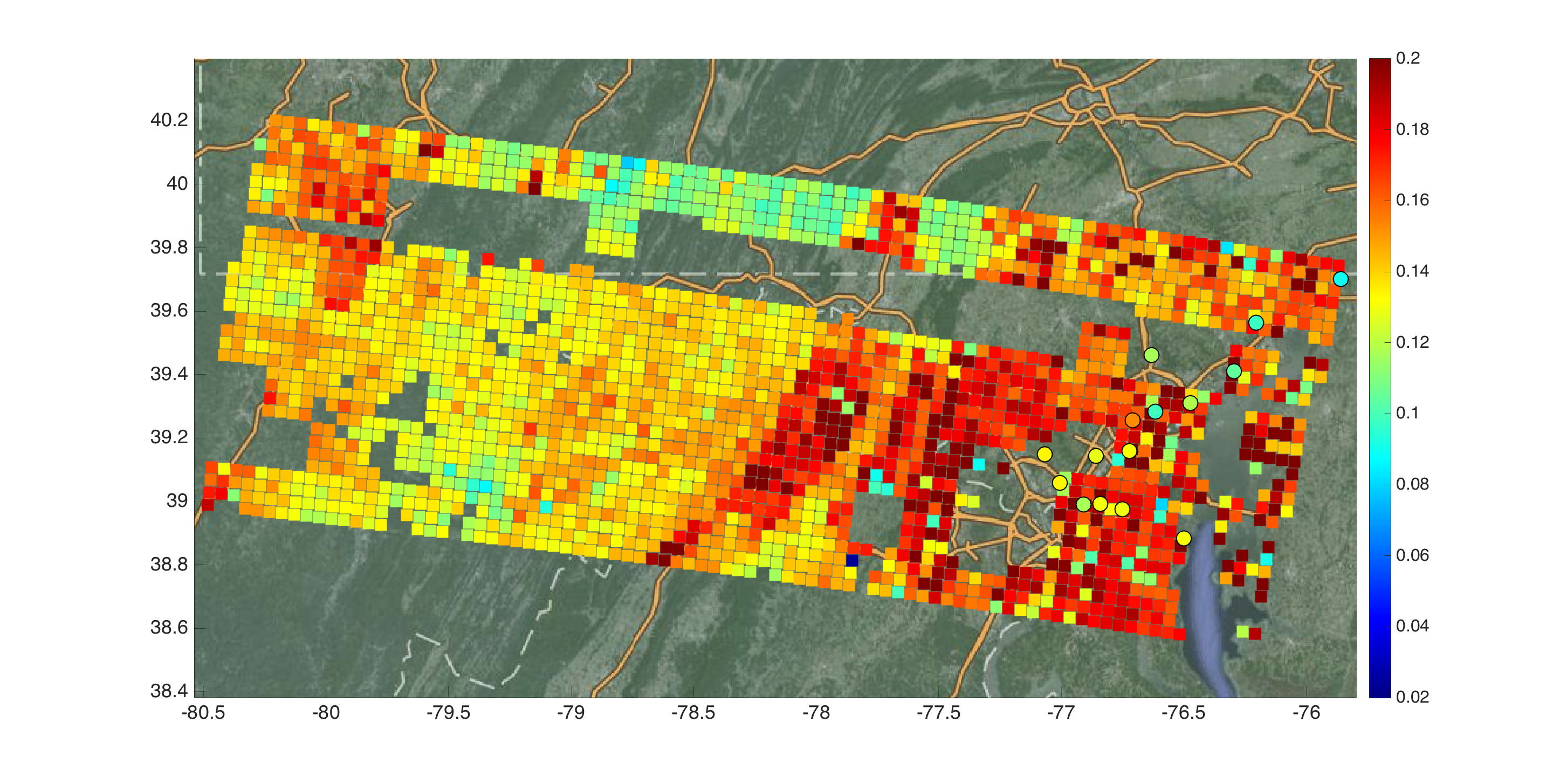}
        \caption{MCMC: $4.4$km $\times$ $4.4$km}
        \label{fig:overlay2}
    \end{subfigure}
    \begin{subfigure}{0.8\textwidth}
        \centering
        \includegraphics[width=\textwidth]{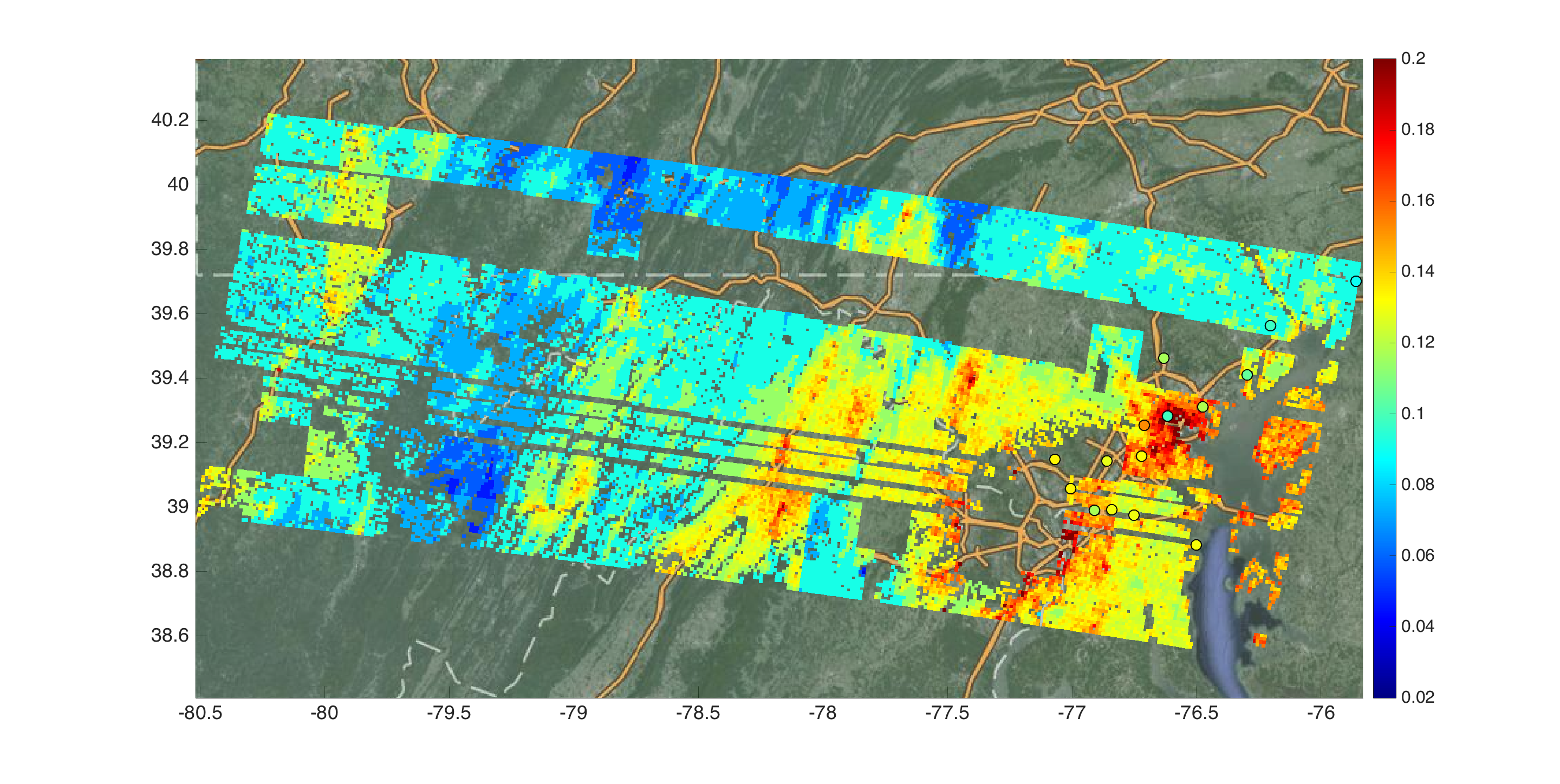}
        \caption{MAP: $1.1$km $\times$ $1.1$km}
        \label{fig:overlay3}
    \end{subfigure}
    \caption{Spatial distribution of AOD retrievals. MAP provides the highest granularity with a tiny fraction of runtime budget.}
    \label{fig:overlay}
\end{figure}

Since MAP is much faster than MCMC, even just given a tiny fraction of the computation time, we are able to perform AOD retrievals using MAP on a much finer spatial grid. Recall that MISR operational algorithm only provides $17.6$~km resolution and MCMC provides $4.4$~km. In this work, we are able to perform AOD retrievals using MAP on $1.1$~km scale with less time. Figure \ref{fig:overlay} shows a comparison among three methods given limited time budge. Obviously within time constraint, MAP with $1.1$~km gives superior spatial details for the retrievals. Color-coded circles represent AERONET data, and are overlaid for qualitative visual check. Obviously, MAP not only produces fine grained AOD retrievals, but the results also agree with AERONET data very well.

\begin{figure}[H]
	\centering
	\includegraphics[width=0.8\textwidth]{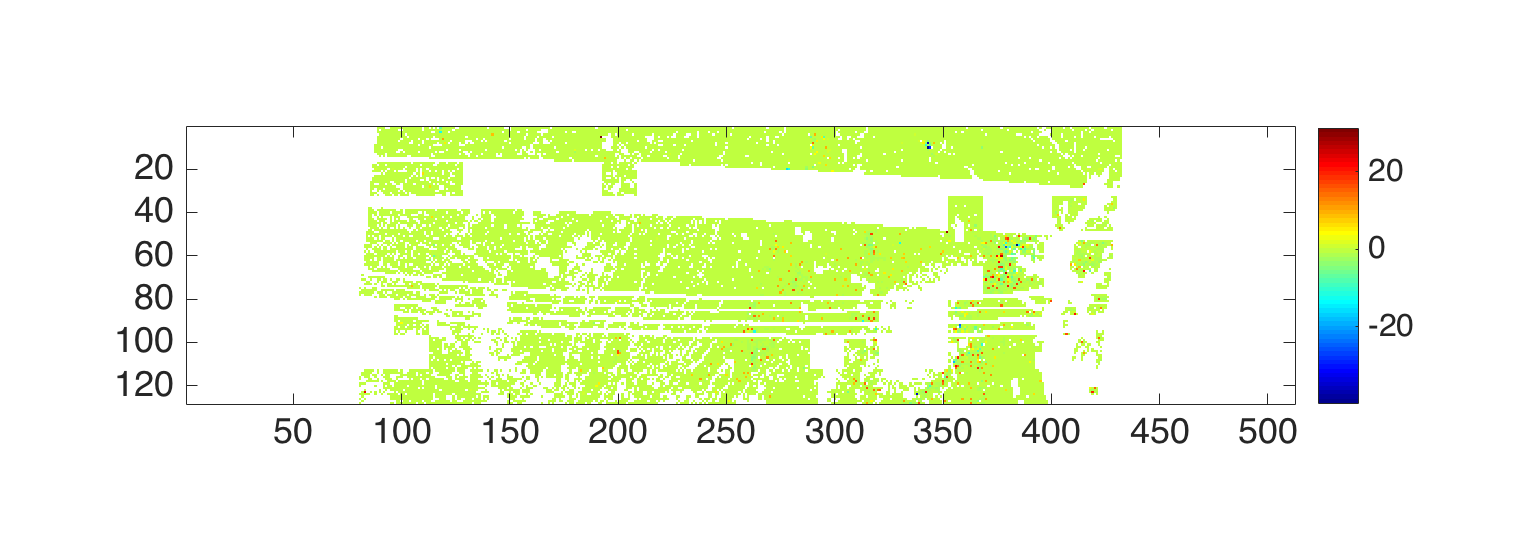}
	\caption{Spatial distribution of 8-core error relative to 1-core (in $\%$). Most pixels have zero error. 8-core parallelization almost achieves the same accuracy as 1-core even with the approximations in each iteration.}
	\label{fig:scaling_cmp}
\end{figure}

As for the algorithm parallelization, 8-core is a reasonable choice as the algorithm takes advantage of the parallelization most effectively. We compare the retrieval results between 1-core and 8-core in Figure \ref{fig:scaling_cmp}, and the difference is almost zero! The results justify our assumption that using last iteration of variables as a surrogate to update the current variables is legitimate.

\begin{figure}[H]
	\centering
	\includegraphics[width=0.8\textwidth]{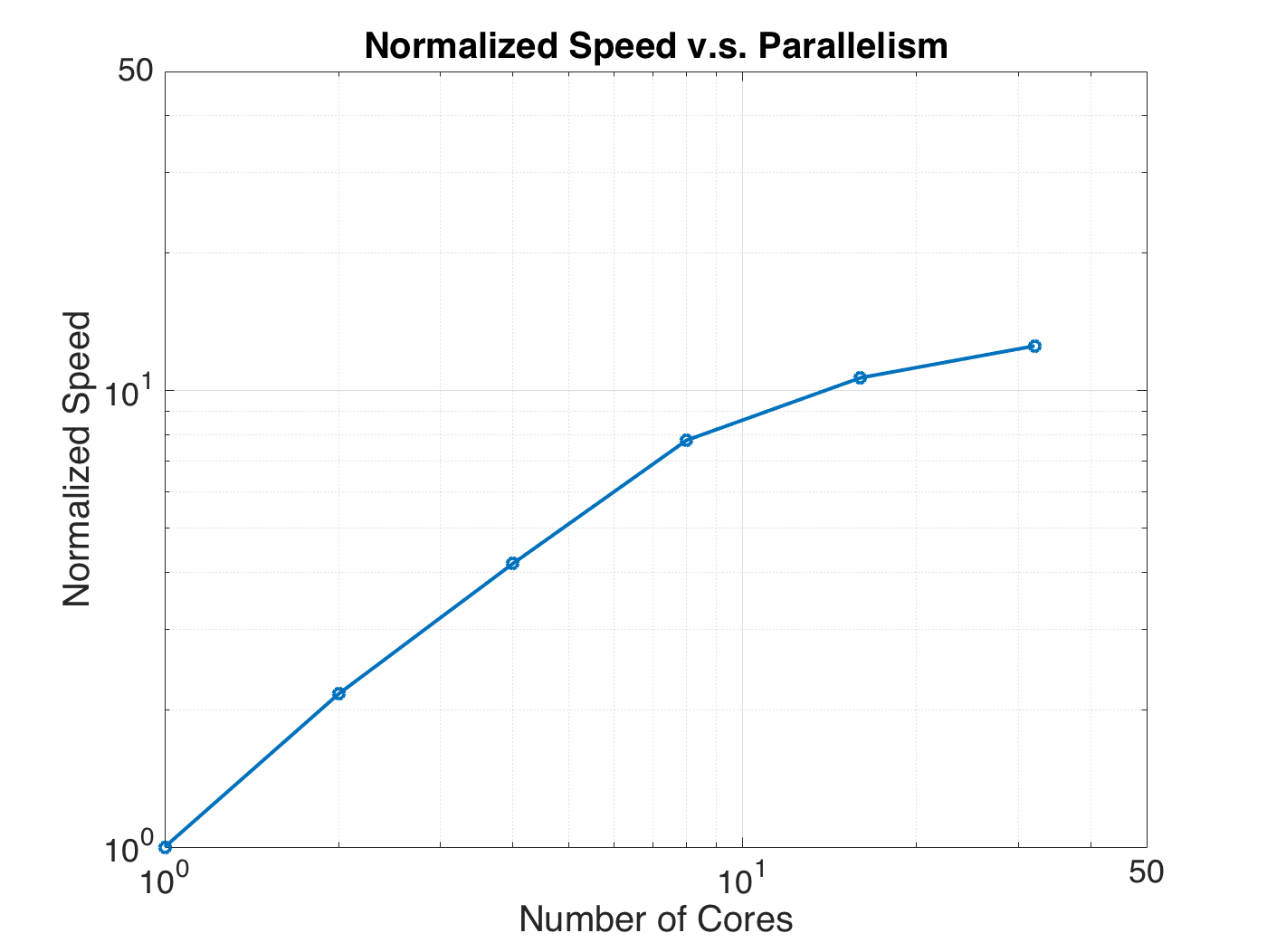}
	\caption{Normalized time speed-up (w.r.t. 1-core) in Spark implementation}
	\label{fig:spark_scaling}
\end{figure}

Figure \ref{fig:spark_scaling} shows that the linear speed-up becomes saturated after 32 cores. We believe this is due to data shuffling overhead. At each iteration when patch-wise update is finished, the algorithm must collect the parameters in different patches to be used for the next round. The more number of cores, the more synchronization is needed, and the speed-up rate ultimately becomes sub-linear.

\section{Discussion}
\label{sec:discussion}

In this section, we make a thorough investigation into the Bayesian model. We want to answer the following three questions:

\begin{itemize}
    \item Why did previous work \cite{taesup} produce overestimation in low AOD levels?
    \item Why does our work continue to use Bayesian framework?
    \item Due to stochasticity in our method, how stable is our retrieval result?
\end{itemize}

We will address these questions by both AERONET data and simulation.

\subsection{Root Cause of Overestimation}

As shown in Figure \ref{fig:scatter} MCMC significantly overestimates aerosol levels on low-AOD end while MAP either doesn't or has only slight overestimation. On the contrary, MISR operational algorithm doesn't have overestimation on low-AOD end at all. On the high-AOD end, Bayesian model with both MCMC and MAP has decent accuracy, but MISR operational algorithm has very poor accuracy - it underestimates aerosol levels significantly. Previous work didn't provide a clear explanation why this would happen. We dig into the data and find the root cause.

\begin{figure}[H]
	\centering
	\includegraphics[width=0.5\textwidth]{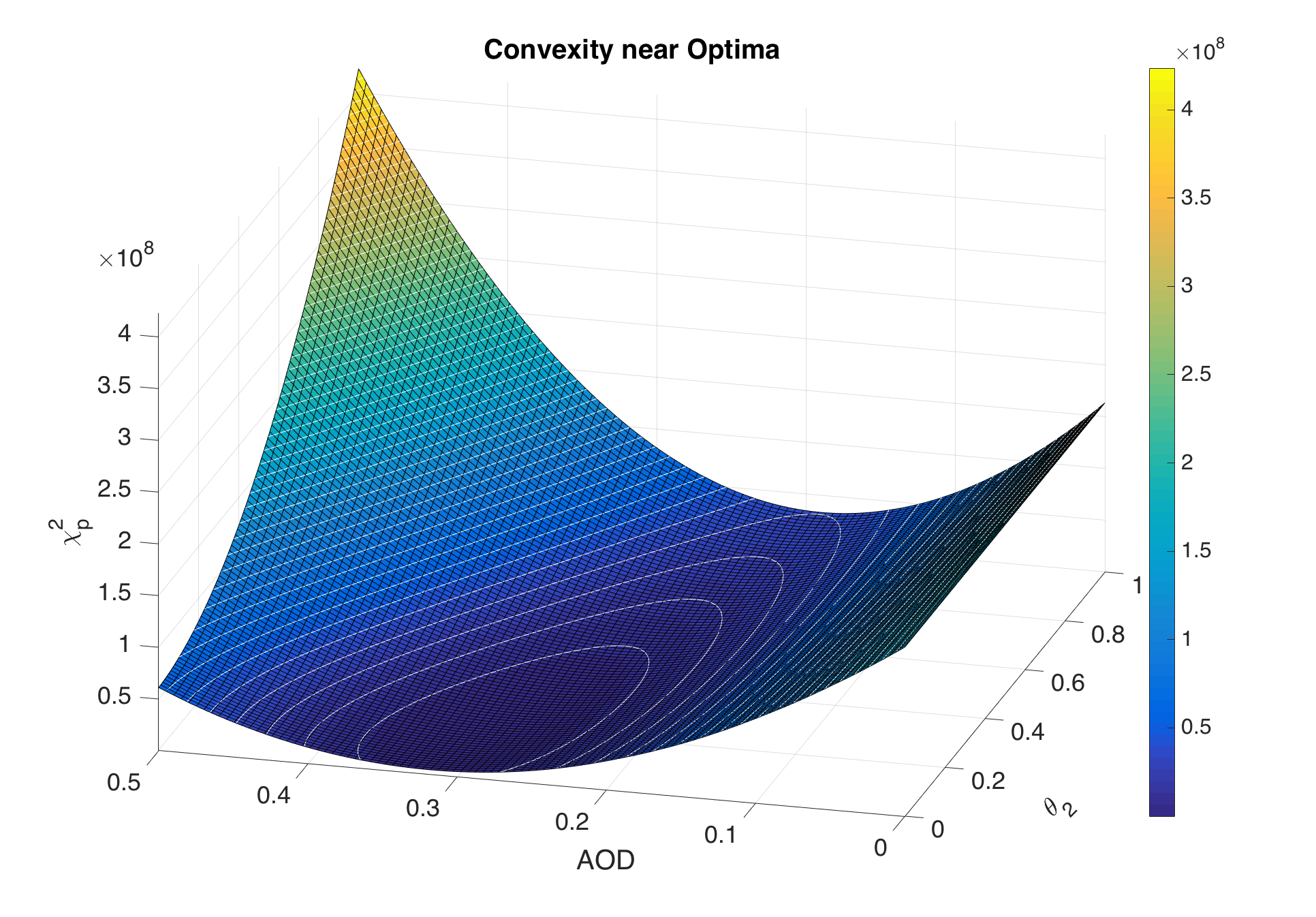}
	\caption{Convexity of negative log-posterior as a function of AOD $\tau$ and mixing percentage $\theta_2$ of aerosol component 2 (spherical non-absorbing aerosols)}
	\label{fig:convexity}
\end{figure}

Retrieving AOD is all about optimization the objective function \eqref{equ:aod_obj}. Note that $\tau_p$ and $\overrightarrow{\theta_p}$ could potentially interact with each other. Inaccurate $\tau_p$ should go hand-in-hand with $\overrightarrow{\theta_p}$.  To confirm $\tau_p$ and $\overrightarrow{\theta_p}$ are not independent of each other, we plot the objective function in Figure \ref{fig:convexity}. The result shows that the two variables are indeed coupled, and in an asymmetrical way - AOD has more sensitivity than with composition mixture.

\begin{figure}[H]
    \centering
    \begin{subfigure}{.3\textwidth}
        \centering
        \includegraphics[width=\textwidth]{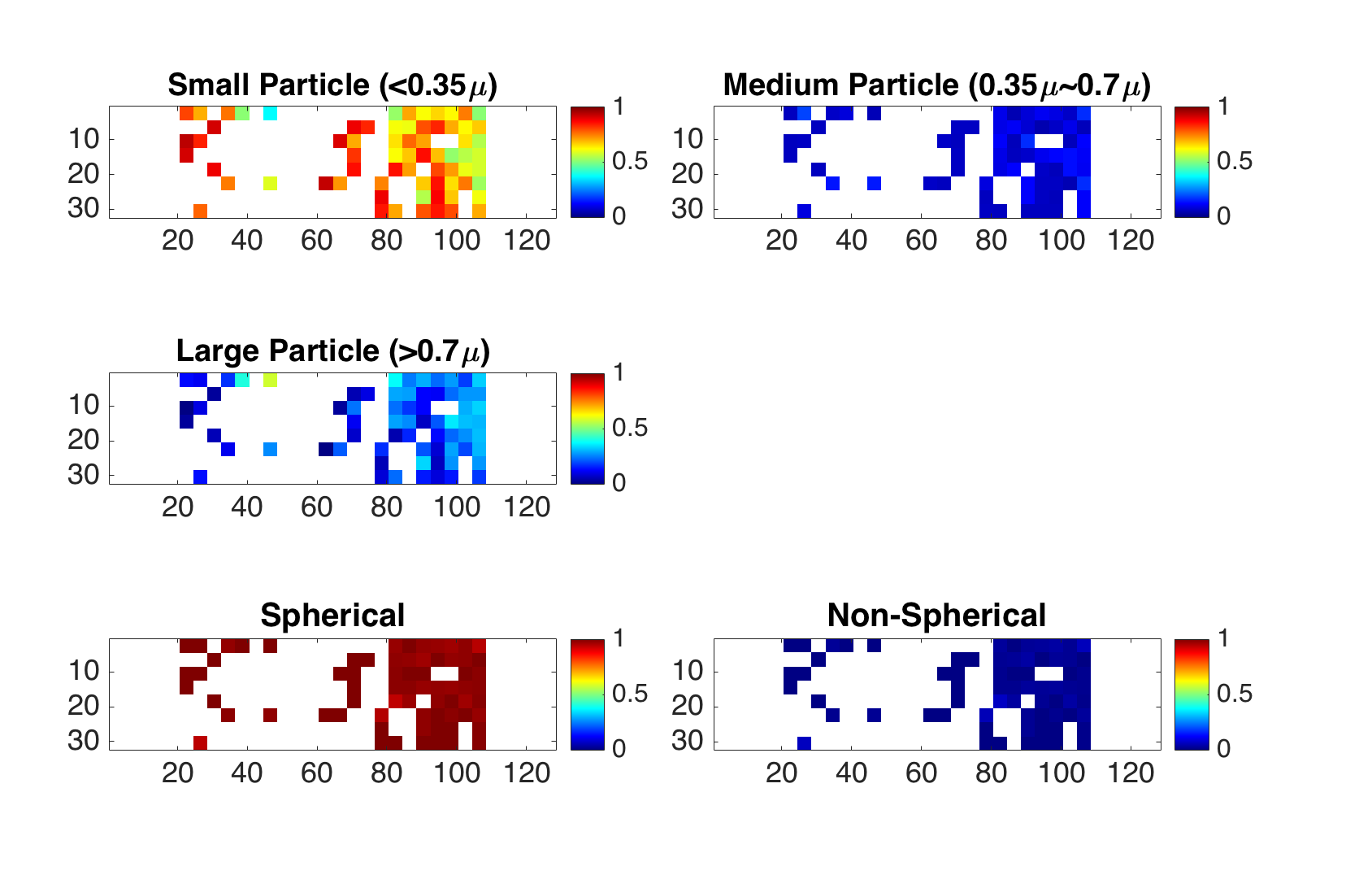}
        \caption{MISR Operational Component Retrieval: small, spherical and non-absorbing dominant}
        \label{fig:composition_cmp1}
    \end{subfigure}
    \begin{subfigure}{.3\textwidth}
        \centering
        \includegraphics[width=\textwidth]{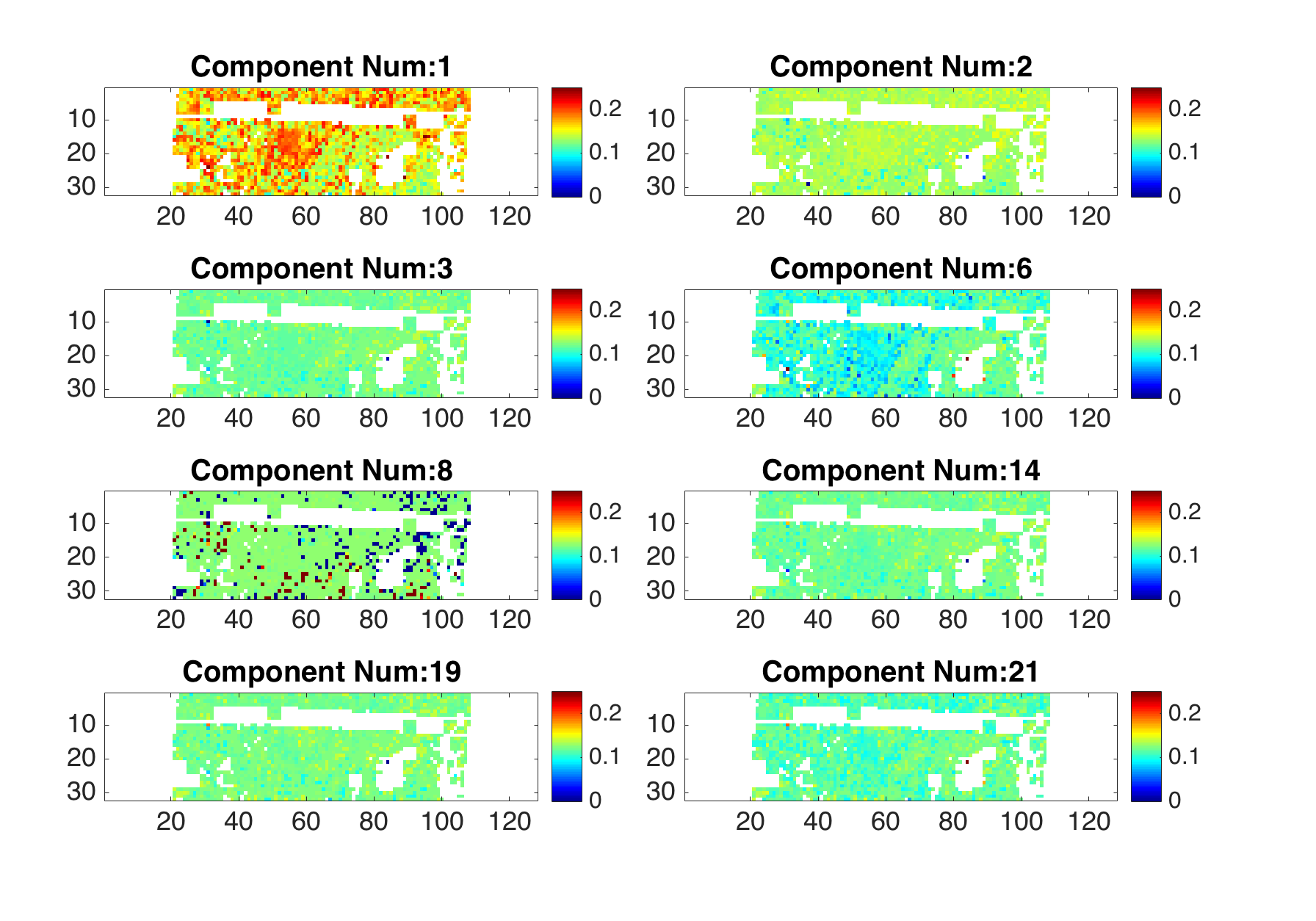}
        \caption{MAP Component Retrieval: small, spherical and non-absorbing dominant}
        \label{fig:composition_cmp2}
    \end{subfigure}
    \begin{subfigure}{.35\textwidth}
        \centering
        \includegraphics[width=\textwidth]{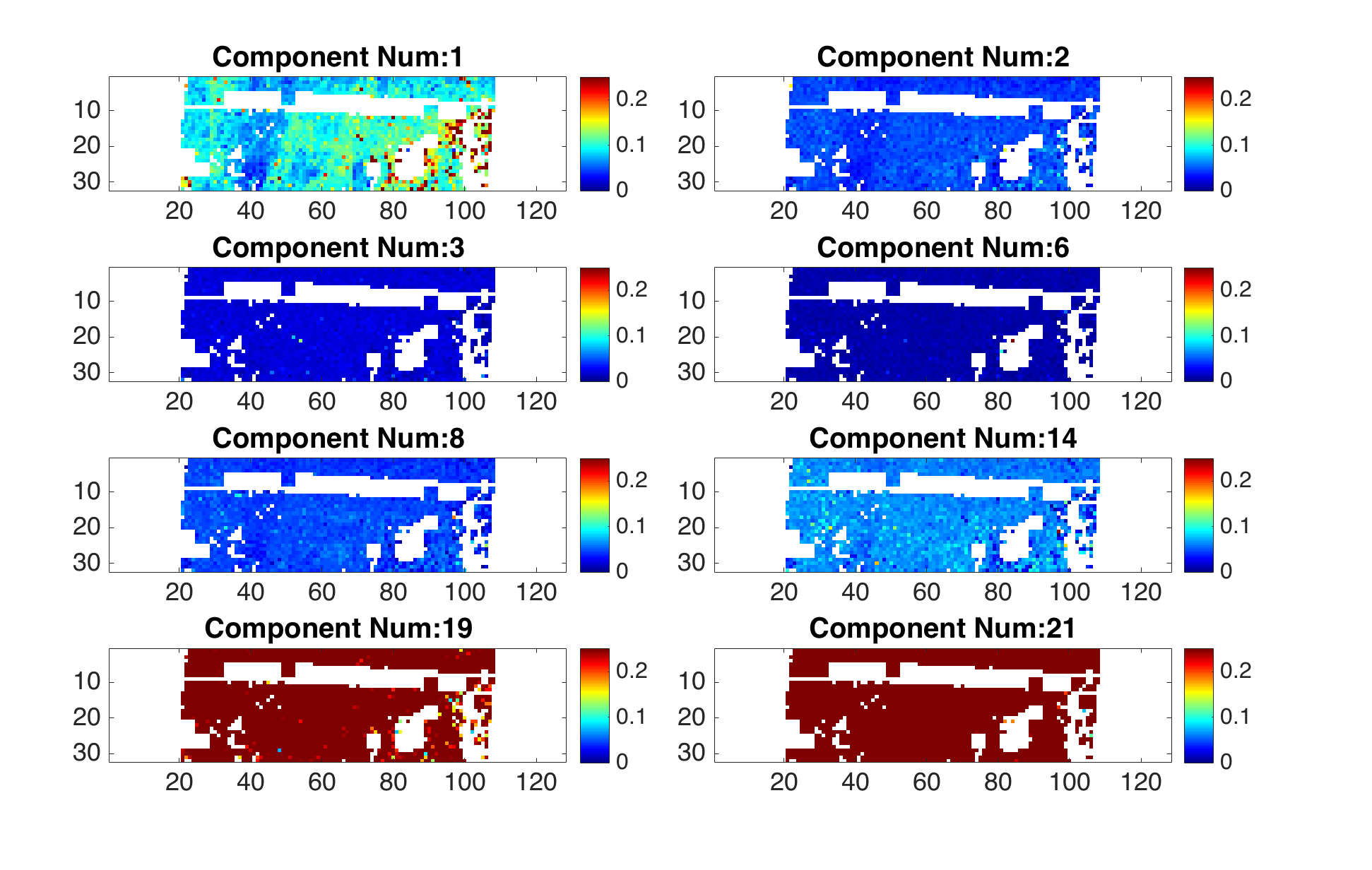}
        \caption{MCMC Component Retrieval: large, absorbing dominant}
        \label{fig:composition_cmp3}
    \end{subfigure}
    \caption{Spatial distribution of aerosol composition retrievals. Different methods lead to very different component choice. Use MISR component retrieval as baseline as ground truth unavailable.}
     \label{fig:composition_cmp}
\end{figure}

Since $\tau_p$ and $\overrightarrow{\theta_p}$ are not independent, we hypothesize that the overestimation in $\tau_p$ is due to incorrect selection of aerosol types in $\overrightarrow{\theta_p}$. Under this hypothesis, overestimation of AOD indicates that too much absorbing types are assumed in the model, causing the corresponding AOD exceedingly high to output the same level of reflectance as MISR would observe. Recall that in Figure \ref{fig:scatter}, AOD retrieval by MCMC near 0.1 has significant overestimation while retrievals by MAP and MISR operational algorithm match with AERONET measurement pretty well. We check the aerosol composition retrievals by MISR operational algorithm, MCMC and MAP respectively and show the results in Figure \ref{fig:composition_cmp}.

Since ground measurement for aerosol composition is unavailable in AERONET measurement, we use MISR operational result as a qualitative baseline to compare with. According to Figure \ref{fig:composition_cmp1}, small ($<0.35\mu m$) and spherical particles should be the dominant aerosol types in the composition retrieval. In Figure \ref{fig:composition_cmp2}, MAP suggests 1 and 2 should be the dominant type while in Figure \ref{fig:composition_cmp3}, MCMC suggests 19 and 21 should be the dominant one. According to Table \ref{tab:misr_comp}, 1 and 2 are small spherical particles while 19 and 21 are absorbing dust (large). The presence of absorbing component in MCMC retrievals explains why its AOD retrieval has severe overestimation. And the composition retrievals exactly match our hypothesis. We further check the AOD near 0.2 and have similar findings.

In the Bayesian framework, our method sets hyper parameter $\overrightarrow{\alpha}$  to be fixed at 1, which allows $\overrightarrow{\theta_p}$ (aerosol component selection) to be unbiased and dense. However in previous work\cite{taesup}, $\overrightarrow{\alpha}$ is given a non-informative prior $\exp\{\sum\limits_{m=1}^M (1-\alpha_m)\}$ where small $\overrightarrow{\alpha}$ is encouraged. In that case, if $\overrightarrow{\theta_p}$ is more likely to be trapped in a local optimum if incorrect sparse aerosol types are chosen by the Dirichlet prior.

\begin{figure}[H]
    \centering
    \begin{subfigure}{.5\textwidth}
        \centering
        \includegraphics[width=\textwidth]{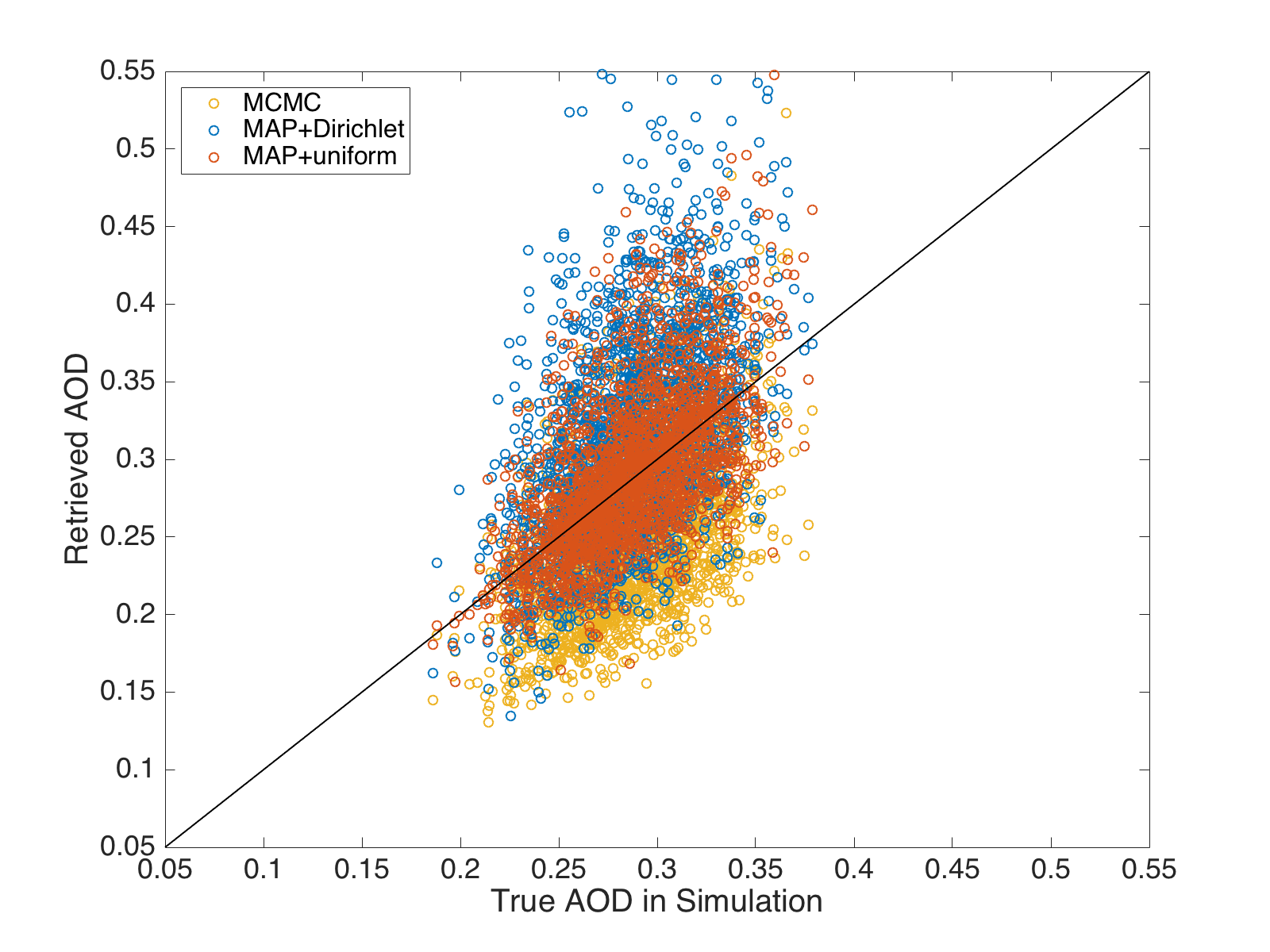}
        \caption{Dirichlet prior is worse than uniform prior in simulated AOD retrieval. (Diagonal line represents perfect result.)}
        \label{fig:cmp00_all_sparse}
    \end{subfigure}%
    \begin{subfigure}{.5\textwidth}
        \centering
        \includegraphics[width=\textwidth]{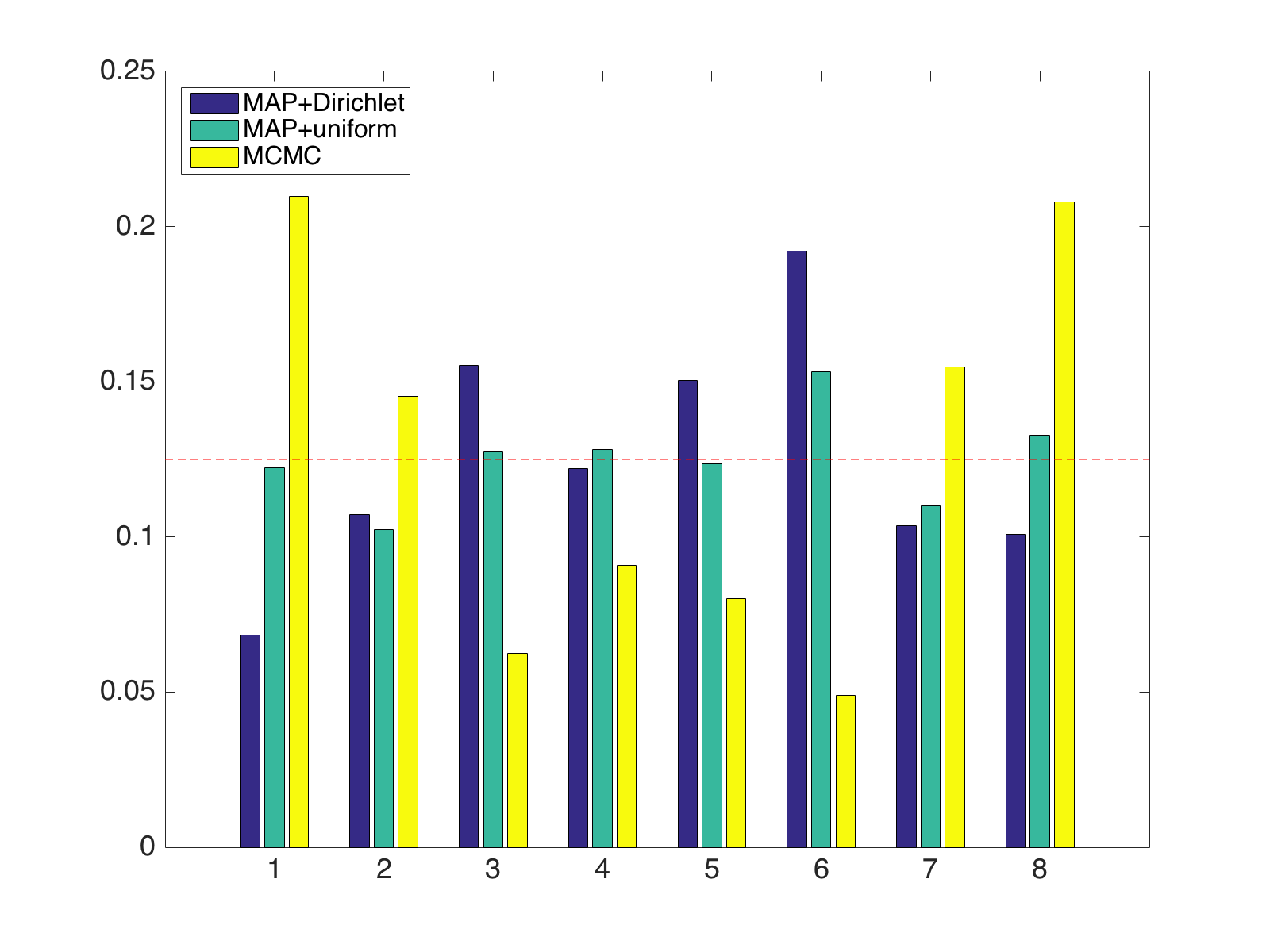}
        \caption{Dirichlet prior is worse than uniform prior in simulated component retrieval. (Red vertical line represents perfect result.)}
        \label{fig:cmp00_all_sparse_kappa}
    \end{subfigure}
    \caption{Uniform prior works better than Dirichlet in MAP}
    \label{fig:cmp00}
\end{figure}

To see whether fixing $\overrightarrow{\alpha}$ at 1 is necessary in general cases, we use simulation data with random component mixture $\overrightarrow{\theta_p} \in (0,1)$ and $\tau_p$. We initialized two versions of $\overrightarrow{\alpha}$: a degenerated prior $Dirichlet(\alpha = 1)$ and a non-degenerated prior $Dirichlet(\alpha = 0.125)$. Figures \ref{fig:cmp00} shows the corresponding results. MAP with Dirichlet prior ends up with very poor AOD retrievals, whereas MAP with uniform prior has excellent ones. The component retrievals confirm that the inaccuracy of AOD retrieval is due to corresponding incorrect composition retrieval.

\subsection{Rationale of Using Bayesian Network}

\begin{figure}[H]
    \centering
    \includegraphics[width=.8\textwidth]{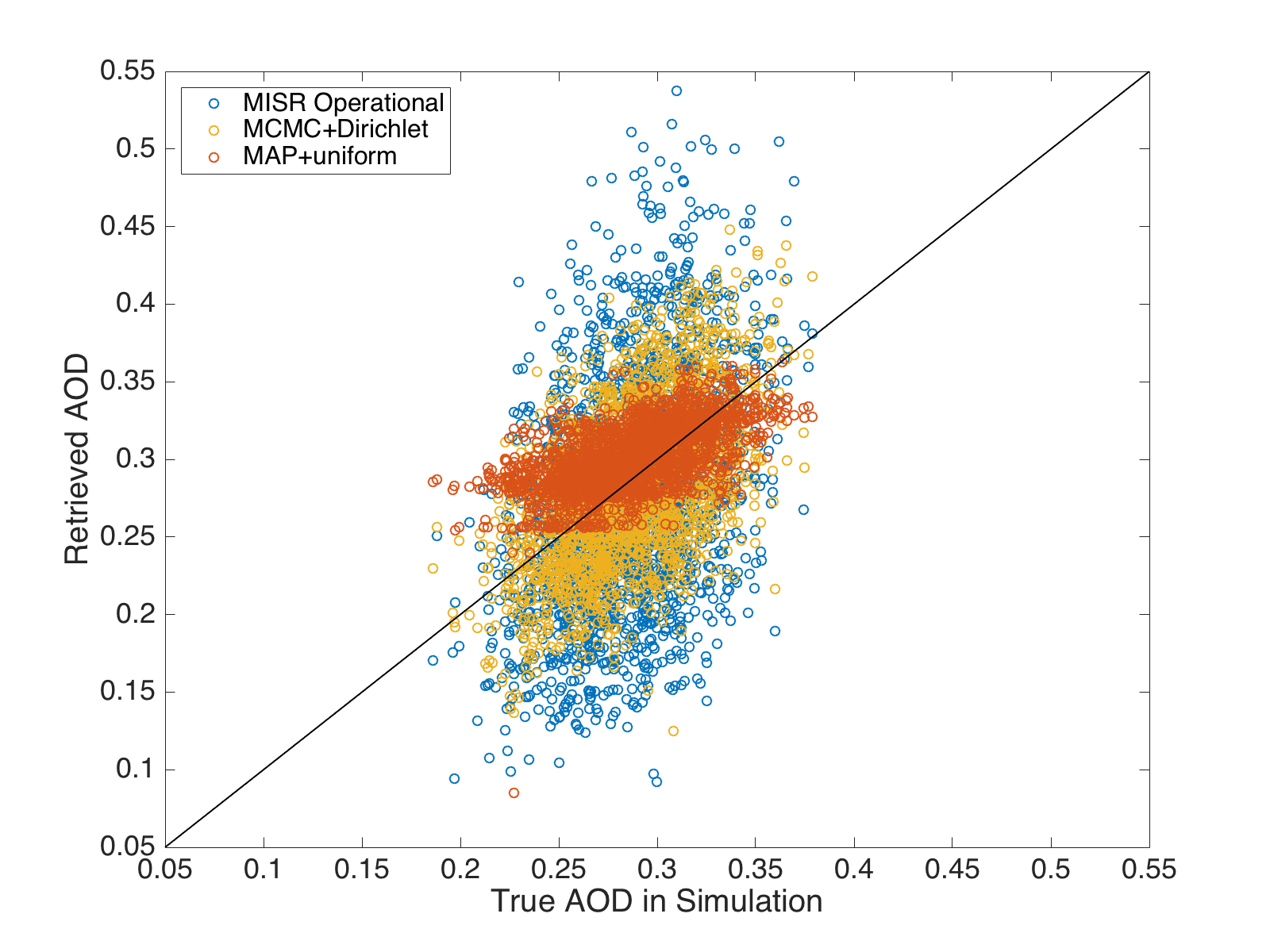}
    \caption{AOD retrievals with simulated image data (50\% noise added). Bayesian methods (MAP and MCMC) show better accuracy than non-Bayesian method (MISR operational algorithm). Diagonal line represents perfect result.}
    \label{fig:cmp50}
\end{figure}

We continue to adopt the same Bayesian framework proposed by \cite{WanJiaYuJia12} because of its immunity to the image noise. This is especially important when we apply this method to high resolution MISR image. To get a qualitative idea about how much the Bayesian model can prevent overfitting in presence of image noise, we did the following experiment. We added 50\% noise to the simulated image data on purpose and then perform AOD retrieval. Figure \ref{fig:cmp50} shows that MISR operational algorithm (non-Bayesian) overfits the data dramatically, whereas MAP (or MCMC) demonstrates excellent result.

\subsection{Stability of AOD Retrievals}
\begin{figure}[H]
    \centering
    \includegraphics[width=.8\textwidth]{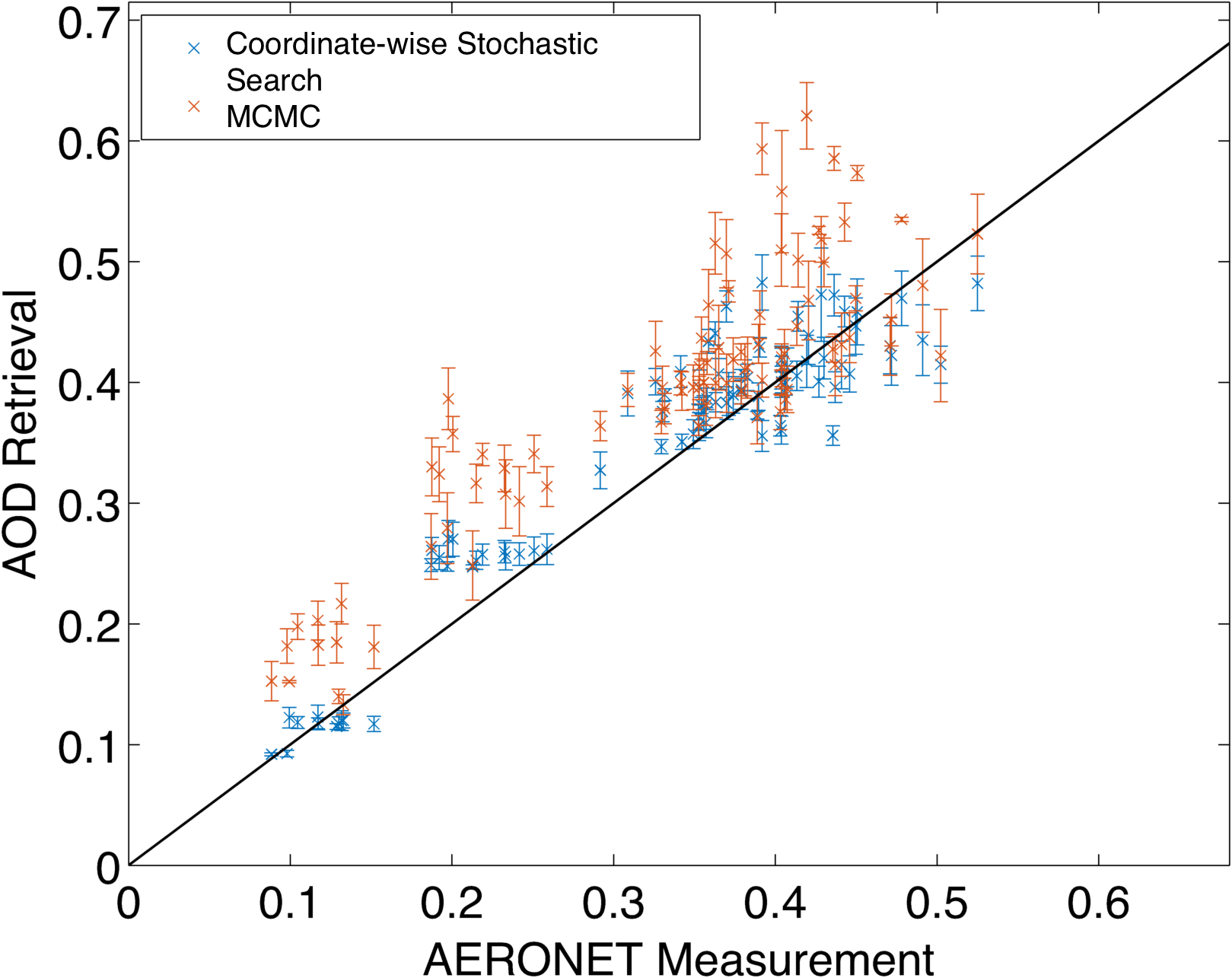}
    \caption{Error Bar for MAP and MCMC. The bounds of stochastic MAP tend to be tighter than those of MCMC.}
    \label{fig:stability_boot2}
\end{figure}

Occasionally people are interested in the stability of the AOD retrievals. Since our method is implemented by coordinate-wise stochastic search, so we can provide such information by different initializations. Figure \ref{fig:stability_boot2} shows an upper bound and lower bound for AOD retrievals based on standard deviation of multiple runs. We can see that the bounds our method provides tend to be tighter than those given by MCMC.

\section{Conclusions and Future Directions}\label{sec:conclusion}

In this paper, we provide an efficient and scalable AOD retrieval algorithm based on the hierarchical Bayesian model proposed in \cite{WanJiaYuJia12}. In particular, we use a stochastic-search based MAP method instead of MCMC to make inference of the parameters, and we are able to parallelize the computation in Spark. Thanks to these efforts, our method is orders of magnitude ($\sim100$ times) faster than existing work \cite{WanJiaYuJia12,MISRAlgTheory} while achieving much finer ($16$ to $256$ times) spatial granularity. Besides, our method is also able to provide stability measure for AOD retrievals.

Further more, we provide an in-depth analysis about why existing work \cite{taesup} tends to overestimate aerosol levels on low-AOD level. We find that too much absorbing component was assumed in their work. We substantiate this finding by checking both AERONET measurement and simulation data. We also advocate that the Bayesian framework is superior to the MISR operational algorithm. The Bayesian method demonstrates excellent reliability in presence of high-level noise. In applications where spatial resolution is high (i.e. 1.1km) and hence image data is noisy, this feature is especially important.

From this work, we also learn that AOD retrievals and aerosol component retrievals go hand in hand. In general when incorrect component is assumed, AOD retrievals will be off accordingly. Although changing the prior of component mixture from Dirichlet to uniform gives us correct component retrievals (and hence correct AOD retrievals) in practice, we cannot theoretically guarantee the method works in general. This is one limitation of our work.

To overcome this limitation, in future work we plan to incorporate the spectral shape of MISR reflectance to retrieve categorical information, since aerosol types are best discerned by spectral information \cite{diner2005}.

Another direction of future work is data fusion with other satellite image data, such as MODIS \cite{kovacs,liu2007,you2015}. Spectrally MISR only has four bands while MODIS has 36. If we can intelligently fuse the data between the two, our capability of  aerosol retrieval would be further enhanced.

\section*{Acknowledgment}

This work was supported by NASA. We thank the great help from JPL MISR science team including Michael Garay, Amy Braverman, Olga Kalashnikova and David Diner. We thank Brent Holben for his effort as the principal investigator in establishing and maintaining AERONET DRAGON sites for the Baltimore-Washington metropolitan region. We also thank Yang Liu's technical explanation of AERONET data. Finally we thank Taesup Moon for all the helpful discussions.

\bibliographystyle{abbrv}
\bibliography{references}

\end{document}